\documentstyle[12pt]{article}
\begin{document}

\title{About the end of the electron spectrum in five--lepton $\mu^+$
decay}
\author{N.P.Merenkov and O.N. Shekhovtsova}
\date{}
\maketitle
\begin{center}

\small{\it{Kharkov Institute of Physics and Technology \\ 61108,
Akademicheskaya 1, Kharkov, Ukraine}}

\end{center} \vspace{0.5cm}
\begin{abstract} The spectrum of very fast electrons in five--lepton decay
$\mu^+\rightarrow e^-e^+e^+\nu\bar\nu$, that is the main background
decay at the study of the muonium--antimuonium conversion in vacuum, is
considered.  The essential decrease of the spectral distribution is
demonstrated when the energy of one positron in this decay is small. Some
arguments for such deacrease for arbitrary positron energies are given.
\end{abstract} \hspace{0.1cm}

\hspace{0.1cm}

1. The search of deviation from Standard Model (SM) and the probe physics
beyond it are the main goals of the current and future experiments in the
elementary particle physics. In this connection, the study of the
spontaneous conversion of muonium ($\mu^+e^-$) into antimuonium
($\mu^-e^+$) is of the great interest nowadays because in
this process the additive lepton family (generation) number would violate
by two units. There are different theoretical models beyond SM where such
violation is permitted [1--5].

In recent experiments [6,7], the observation of muonium atom in
vacuum have been used to investigate the conversion process. The conversion
event would manifest itself by the registration of the fast electron that
appears due to standard decay of $\mu^-$ from antimuonium atom
\begin{equation}\label{1}
e^-\mu^+\rightarrow e^+\mu^- \rightarrow e^+ + e^- + \nu + \bar{\nu} \ .
\end{equation}

The main electrodynamical background for such events in vacuum arises due
to five--lepton decay of $\mu^+$ in muonium
\begin{equation}\label{2}
\mu^+(p) \rightarrow e^-(p_3)+e^+(p_1)+e^+(p_2) + \nu\bar{\nu}(q)\ .
\end{equation}

It is well known that in the case when the energies of both positrons in
decay (2) are small enough, the energy distribution of the electrons
is strong suppressed at the end of their spectrum. The form of the
spectrum at these conditions is [8,9]
\begin{equation}\label{3} \frac{d\Gamma}{\Gamma_0\
dy}=\frac{\alpha^2}{\pi^2}(1-y)^2 F(L,l) \ , \ \ L=\ln\frac{M^2}{m^2}\ , \
\ l=\ln(1-y)\ , \ \ \Gamma_0 = \frac{G^2M^5}{192\pi^3}\ , \end{equation}
$$ 1-y \ll 1,  \ \ x_1 + x_2 < 1-y , \ \ y
=\frac{2\varepsilon_-}{M}\ , \ \ x_{1,2} = \frac{2\varepsilon_{1,2}}{M} \
, $$ where $M(m)$ is the muon (electron) mass, $\varepsilon_-, \
\varepsilon_{1,2}$ are energies of the electron and positrons,
respectively, and $F(L,l)$ is the known function (see below). The
additional smalness $(1-y)^2$ arises because the trivial phase space
factor of positron energy fractions: $\Delta x_1\Delta x_2\approx (1-y)^2.$

Just this additional smallness of the probability of the background decay
(2) makes the selection events near the end--point of the electron
spectrum in the process (1) very attractive to observe the
muonium--antimuonium conversion.

In Ref. [6] it is suggested that in the general case, where the
positron energy fractions $x_1$ and $x_2$ can be arbitrary possible, the
form (3) of the differential width breaks down and the small factor
$(1-y)^2$ disappears. In present work we want to explain that this factor
remains independent on values of $x_1$ and $x_2.$ The physical reason for
this assertion is the decrease of the angular phase space of the positron
with the large energy. Namely, if the electron in decay (2) carries away
the energy fraction $y$ such that $(1-y)\ll 1 $ and positron with
4--momentum $p_1\ \ (p_2)$ has the energy fraction $x_1\ \ (x_2) \gg
(1-y)$ then it fly just in the opposite direction respect to the energetic
electron one and $\Delta c_1 \Delta c_2\sim (1-y),$ where $c_{1,2} =
\cos{\widehat{\vec p_{1,2}\vec p_3}}.$ We calculate analytically the
function $F(L,l)$ for the case when the energy of one positron is smaller
than $1-y$ and the energy of the other one is arbitrary. When calculating
we neglect with terms of the order of $(1-y)$ in this function.

2. In our calculations we do not take into account the
indentity of positrons, because the coresponding effects consist no
more than five per cent in a general case (in accordance with Monte Carlo
calculations [10]) and trend to decrease in the considered here case
when $1-y \ll 1$ [9].  Besides, we use the relativistic
approximation and neglect the electron mass always where it is possible.
We start from the differential width of the five--lepton decay (2) in the
following form \begin{equation}\label{4} \frac{d\Gamma}{\Gamma_0 d\ y} =
\frac{\alpha^2}{4\pi^2}Rx_1x_2dx_1dx_2dc_1\frac{d\Omega_2}{2\pi}\ ,
\end{equation}
where in the used approximation the quantity $R$ can be written as a
contraction of two tensors [9]
\begin{equation}\label{5}
R=\bigl[a_1\widetilde g_{\mu\nu} +2a_2\tilde
p_{2\mu}\tilde p_{2\nu}+2a_3\tilde p_{\mu}\tilde p_{\nu} +2a_4(\tilde
p\tilde
p_2)_{\mu\nu}\bigr]\bigl[(p_1p_3)_{\mu\nu}-\frac{k^2}{2}g_{\mu\nu}\bigr] \
,
\end{equation}
$$k^2=(p_1+p_3)^2\ , \ \ \widetilde g_{\mu\nu} =
g_{\mu\nu}-\frac{k_{\mu}k_{\nu}}{k^2}\ , \ \ \tilde p_{\mu} =
p_{\mu}-\frac{pk}{k^2}k_{\mu}\ , \ \ (ab)_{\mu\nu} =
a_{\mu}b_{\nu}+a_{\nu}b_{\mu}\ , $$
and quantities $a_i$ on the right side
of Eq.~(5) are given in Appendix 1 of Ref. [9] (Preprint).

The constraints on the possible angles and energy fractions of
positrons can be obtained from the condition on the invarinant mass of
neutrinos: this last have to be positive
\begin{equation}\label{6}
1-y-x_1-x_2+\frac{x_1y}{2}(1-c_1)+\frac{x_2y}{2}(1-c_2)+\frac{x_1x_2}
{2}(1-c_{12}) > 0\ ,
\end{equation}
where $c_{12} =\cos{\widehat{\vec p_1\vec p_2}}.$

To simplify calculations we divide the positron energy fractions phase
space by four kinematical regions, where restrictions on the
positron energy fractions, positron angles and form of R on the right
side of Eq.~(3) are different. In the first region the condition
$$1-x_1-x_2-y \geq 0$$ is satisfied. One can see from inequality (6) that
in this case all angles for positrons are permitted.  The quantity $R$ in
this region is very simple \begin{equation}\label{7} R_1 =
\frac{M^4}{2}\Bigl(\frac{1}{k^2B}-\frac{1}{B^2}\Bigr)\ , \ \
B=(p_1+p_2+p_3)^2\ .  \end{equation} In the framework of chosen accuracy we
can use $k^2+2p_2p_3$ for $B$ in $R_1$ because the quantity $p_1p_2$ is
always negligible in this region. Then the integration $R_1$ over the
phase space of positrons gives well known result obtained in [8]
(see also [9]) \begin{equation}\label{region 1}
F_1(L,l)=\frac{1}{8}L^2+L\bigl(-1+\frac{1}{2}l\bigr)+\frac{1}{2}l^2-2l+3-
\frac{\pi^2}{12} +O(1-y) \ .
\end{equation}

In the second region the energy fractions of positrons satisfies
inequalities
\begin{equation}\label{9}
1-y-x_2<x_1<1-y, \ \ 0 <x_2<1-y\ .
\end{equation}
Because of the smallness both $x_1$ and $x_2$ in this region we can omit
the last term in condition (6) and obtain the constraints on the $c_1$ and
$c_2$ in the form
\begin{equation}\label{10}
-1<c_2<1, \ \ -1<c_1<1+\frac{2(1-x_1-x_2-y)}{x_1y} \ ,
\end{equation}
and
\begin{equation}\label{11}
-1<c_2<1+\frac{2(1-x_1-x_2-y)}{x_2y}+\frac{x_1}{x_2}(1-c_1)\ , \ \
1>c_1>1+\frac{2(1-x_1-x_2-y)}{x_1y}\ .
\end{equation}

With the chosen accuracy the expression for $R$ in the second
region coincides with $R_1.$ The angular integration of separate terms in
$R_1$ over regions (10) and (11) gives
$$\int\limits_{(2)}dc_1dc_2\frac{M^4}{k^2B}=\frac{4}{x_1x_2}\bigl[-L\ln\frac
{x_1+x_2+y-1}{yx_2}-Li_2\bigl(-\frac{x_1}{x_2}\bigr)-\frac{1}{2}\ln^2\frac
{x_1(x_1+x_2+y-1)}{y}$$
\begin{equation}\label{12}
+2\ln x_1\ln x_2\bigr] \ , \ \ \int\limits_{(2)}dc_1dc_2\frac{M^4}{B^2}=
-\frac{4}{x_1x_2}\ln\frac{(x_1+x_2)(x_1+x_2+y-1)}{x_1x_2}\ ,
\end{equation}
where the form of $B$ on the left side of relations (12) is the same as in
the first kinematical region.

The further integration respect to energy
fractions of the positrons defines the contribution of the second
kinematical region (see inequalities (9)-(11)) into function $F(L,l)$
\begin{equation}\label{13} F_2(L,l)=\frac{1}{4}L+\frac{1}{2}l+\ln
2-2+\frac{\pi^2}{12}\ .  \end{equation}

The restrictions on the energy fractions in the third region are as
follows
$$ 0<x_1<1-y\ , \ \ 1-y < x_2 < 1\ . $$
Note that the condition on the energies of the visible particles in
decay (2) gives for the upper limit of $x_2$ the quantity $2-y-x_1,$
which differs from $1$ by value of the order $1-y.$ The accounting of that
difference leads to contribution of the order $1-y$ in $F(L,l)$ which
is beyond our accuracy.

Let us analyze the condition (6) in the third kinematic region.  If
the values $x_2$ are near $(1-y)$ one can, as before, neglect the last term
in (6) and obtain for $c_2$ the same constraints as in the first
inequality in (11).  The extension them on large values of $x_2\gg(1-y)$
means that corresponding values of $c_2$ are near $-1$ with $\Delta
c_2\sim (1-y).$ The physical content of this circumstance is very
transparent: in event with large--energy electron $(y\approx 1)$ the large
energy positron $(x_2\gg 1-y)$ must fly in the opposite direction respect
to the direction of the electron 3--momentum. Contrary the conservation of
3--momentum in decay (2) would be violated.

Therfore, we can correct the
above restriction taking into account the last term in (6) at $c_2=-1.$
This way we derive the constraints on the $c_2$ and $c_1$ in the third
region \begin{equation}\label{14,region 3} -1 < c_2 < -1
+\frac{2(1-y-x_1)(1-x_2)}{x_2y}+\frac{x_1(y-x_2)}{x_2y}(1-c_1)\ , \ \ -1 <
c_1 < 1 \ .  \end{equation}

The quantity $R$ in the third region reads
\begin{equation}\label{15}
R_3=\frac{M^4}{B^2}\bigl(-\frac{1}{2}+\frac{x_2}{2}+2x_2^2\bigr)+\frac
{M^2}{k^2}\big(-1+2x_2+6x_2^2)+\frac{a_{23}}{k^2}\big(-\frac{1}{2}-7x_2\big)+
\end{equation}
$$\frac{3a_{23}^2}{M^2k^2}+\frac{M^4}{k^2B}\bigl(\frac{1}{2}-x_2^2-2x_2^3
\bigr)\ , \ \ a_{23} = 2p_2p_3 \ . $$
In accordance with our prescription in this region we have to take term
$2(p_1p_2)$, that enters in $B,$ at $c_2=-1.$ Such procedure
leads to $$\frac{M^2}{B}=
\frac{2}{x_2}\bigl(1-c_2+2x_1+\frac{x_1}{x_2}(y-x_2)(1-c_1)\big)^{-1}$$
on the right side of Eq.~(15).

The list of necessary angular integrals in $R_3$ reads
$$\int\limits_{(3)}dc_1dc_2\frac{M^4}{B^2} =
\frac{4}{x_2x_1}\big(\frac{x_1}{x_1+x_2+y-1}-
\ln\frac{x_1+x_2}{x_2}\bigr)\ , $$
$$\int\limits_{(3)}dc_1dc_2\frac{M^2}{k^2}=\int\limits_{(3)}\frac{dc_1dc_2}
{x_2}\frac{a_{23}}{k^2}=\int\limits_{(3)}\frac{dc_1dc_2}{x_2^2}\frac
{a_{23}^2}{k^2}=
\frac{4(1-x_2)}{x_2x_1}\bigl(x_1+(1-y-x_1)(L+2\ln x_1)\bigr)\ ,
$$
\begin{equation}\label{16}
\int\limits_{(3)}dc_1dc_2\frac{M^4}{k^2B}=\frac{4}{x_1x_2}\bigl[(L+2\ln
x_1)(\ln\frac{x_2(y+x_1)}{x_1+x_2+y-1}-x_2\ln\frac{x_1+x_2}{x_2}-Li_2
\bigl(-\frac{x_1}{x_2}\bigr)\bigr]\ .
\end{equation}
When writing these integrals we neglect all terms which lead to terms of
the order $(1-y)^3$ in the electron spectrum.

Using Eqs.~(16) and integrating over the posintron energy fractions we
derive the contribution of the third kinematical region into function
F(L,l) in the form \begin{equation}\label{17}
F_3(L,l)=-\frac{1}{3}L-\frac{1}{4}Ll-\frac{1}{2}l^2-\frac{1}{6}l
+\frac{31}{24} -\frac{\pi^2}{24}-\ln 2 \ .
\end{equation}

In the fourth kinematical region
$$ 0 < x_2 < 1-y \ , \ \  1-y < x_1 < 1\ . $$
The restrictions on the $c_1$ and $c_2$ in this region can be obtained
from (14) by the simple substitution
$$x_1\longleftrightarrow x_2\ , \ \ c_1\longleftrightarrow c_2 \ $$
because of obvious symmetry of the positron phase space in the third and
fourth regions.

The expression for $R$ in this region has the following form
\begin{equation}\label{18}
R_4=\frac{M^4}{B^2}\bigl(-\frac{1}{2}+x_1\bigr)(1+x_1) +\frac{M^4}{k^2B}
\bigl(\frac{1}{2}-\frac{x_1}{2}-x_1^2
-\frac{3x_1}{1+x_1}\bigr)+\frac{M^4}{k^4}\frac{x_1(1-2x_1)}{1+x_1}
 \ .
\end{equation}

The angular integrals in considered case are
$$\int\limits_{(4)}dc_1dc_2\frac{M^4}{B^2}=\frac{4}{x_1x_2}\bigl(\frac
{x_2}{x_1+x_2+y-1}-\ln\frac{x_1+x_2}{x_1}\bigr)\ , $$
$$\int\limits_{(4)}dc_1dc_2 \frac{M^4}{k^4}=\frac{4}{x_1^2}\bigl(-1+\ln
\frac{x_1+x_2+y-1}{x_1+y-1}\bigr) , $$
$$\int\limits_{(4)}dc_1dc_2\frac{M^4}{k^2B}=\frac{4}{x_1x_2}\bigl
[Li_2\bigl(\frac{x_2}{x_1+x_2+y-1)}\bigr)+Li_2\bigl(-\frac{x_2}{x_1}
\bigr)-$$
\begin{equation}\label{19}
Li_2\bigl(\frac{x_1x_2}{x_1+x_2+y-1}\bigr)-Li_2(-x_2)\bigr]\ .
\end{equation}

Using expression for $R_4$ and angular integrals (19) we perform the
integration respect to the positron energy fractions in fourth region and
derive \begin{equation}\label{20} F_4(L,l)=\frac{3}{4} -\ln 2\ .
\end{equation}

Thus, the spectrum of very fast electrons in five--lepton decay (2)
(provided the energy fraction of one positron is smaller then $1-y$ and
the energy fraction of other one is arbitrary) is defined as a sum of
contributions  of considered above four kinematical regions and can be
written in the following form
\begin{equation}\label{21}
\frac{d\Gamma}{\Gamma_0d\
y}=\frac{\alpha^2}{\pi^2}(1-y)^2\bigl[\frac{1}{8}L^2+\bigl(\frac{1}{4}l-
\frac{13}{12}\bigr)L-\frac{5}{3}l-\ln 2-\frac{\pi^2}{24}+\frac{73}{24}
\bigr]\  .
\end{equation}
Note that terms containing $L^2$ and $L$ in our final result coincide with
those computed in [9], where collinear and semicollinear
kinematics of the five--lepton $\mu$ decay has been investigated.

Here we considered the case when the energy
of only one (from two) positron in decay (2) is small.  But we sure that
factor $(1-y)^2$ in the fast electron spectrum would appear at arbitrary
positron energies because of essential squeeze of the angular phase space
of the positrons along direction opposite to the electron 3--momentum, if
their energies became large enough. The effect of such squeeze we observed
in our analytical calculations. Therefore, we conclude that at study of
the muonium--antimuionium conversion in vacuum by the observation of
the very fast electron near its maximum energy, the probability of the main
background decay always is very small. For example, in accordance with
our estimations it consists for about $1.24\cdot 10^{-7}\Gamma_0$ if
$y=0.9$ and is on the essential decrease when the electron energy goes on
to increase. \\

\hspace{0.2cm}
{\large {\bf References}}
\begin{enumerate}
\item P. Herczeg and R.N. Mohapatra, Phys. Rev. Lett. {\bf 69},
2475 (1992).
\item A.Halprin, Phys. Rev. Lett. {\bf 48}, 1313 (1982).
\item G.G. Wong and W.S. Hou, Phys. Rev. D {\bf 50}, R2962 (1994);
W.S. Hou and G.G. Wong, Phys. Rev. D {\bf 53}, 1537 (1996).
\item A. Halprin and A. Masiero, Phys. Rev. D {\bf 48}, 2987,
(1993).
\item H. Fujii, Y. Minura, K. Sasaki and T. Sasaki, Phys. Rev. D
{\bf 49}, 559 (1994).
\item V.A. Gordeev et al., JETP Lett. {\bf 59},
589 (1994); Yad. Fiz. {\bf 60}, 1291 (1997).
\item L. Williman et al., Phys. Rev. Lett. {\bf 82}, 49 (1999).
\item E.G. Drukarev et al., Preprint LNPI--1317 (1987); E.G.
Drukarev and V.A. Gordeev, Preprint LNPI--1588 (1990).
\item A.B. Arbuzov, E.A.  Kuraev, N.P.  Merenkov, JINR Preprint
E4--93--87, Dubna 1993; Pis'ma Zhur. Exp. Teor. Fiz. {\bf 57}, 746 (1993).
\item D.Yu.  Bardin et al., JINR Preprint E2--5904, Dubna, 1971.
\end{enumerate}

\end{document}